# "It is there, and you need it, so why do you not use it?" Achieving better adoption of AI systems by domain experts, in the case study of natural science research

Auste Simkute

University of Edinburgh, Edinburgh, United Kingdom, a.simkute@sms.ed.ac.uk

Ewa Luger

University of Edinburgh, Edinburgh, United Kingdom, ewa.luger@ed.ac.uk

Michael Evans

BBC Research and Development, Salford, Greater Manchester, United Kingdom, michael.evans@bbc.co.uk

Rhianne Jones

BBC Research and Development, Salford, Greater Manchester, United Kingdom, rhia.jones@bbc.co.uk

Artificial Intelligence (AI) is becoming ubiquitous in domains such as medicine and natural science research. However, when AI systems are implemented in practice, domain experts often refuse them. Low acceptance hinders effective human-AI collaboration, even when it is essential for progress. In natural science research, scientists' ineffective use of AI-enabled systems can impede them from analysing their data and advancing their research. We conducted an ethnographically informed study of 10 in-depth interviews with AI practitioners and natural scientists at the organisation facing low adoption of algorithmic systems. Results were consolidated into recommendations for better AI adoption: i) actively supporting experts during the initial stages of system use, ii) communicating the capabilities of a system in a user-relevant way, and iii) following predefined collaboration rules. We discuss the broader implications of our findings and expand on how our proposed requirements could support practitioners and experts across domains.

# 1 INTRODUCTION

Artificial Intelligence (AI) is increasingly important in domains such as public sector administration [22], medicine [54], and life sciences [64,84]. AI is intended to automate mundane, repetitive tasks and focus human expertise on higher-level and creative activities [34,68,82,83]. In many contexts, AI is a crucial solution to coping with the vast data available due to technological advances [26]. Complex AI systems and the availability of biological data have already contributed to significant achievements in fields such as Alzheimer's research [84], drug discovery [53] and disease diagnostics [40]. However, without AI being appropriately implemented, there is a risk of "bottleneck" situations in domains such as the natural sciences [65]. If scientists are constrained to manual data processing, valuable data is left unused, and significant scientific discoveries are slowed down or halted [47]. Domain experts also acknowledge the benefits AI systems offer [79], yet in practice, they remain reluctant to include AI in their workflows [16], they use AI systems ineffectively [33,46,76], or their performance worsens with the support of AI [15,16,24,81].

The Human-Computer Interaction (HCI) community has approached this issue of AI resistance by proposing more usable, accurate, and explainable technologies [30,59,60]. Increasing efforts have been directed at understanding how model transparency influences the adoption of AI systems in practical settings, for example, by studying physicians' interactions with AI [14]. These efforts are undoubtedly needed to develop ethical and user-centred technologies. However, domain experts prematurely reject and mismanage AI systems, even if they are explainable and outperform them [15,16]. However, despite the growing user-centred research focus, in practice, experts' needs continue to be misinterpreted in the design of these systems, leaving them frustrated with technologies that do not fit their needs, knowledge, and workflows [12,66]. We argue that more fundamental and underexplored barriers exist between experts and AI tools designed to support them. First, there is a lack of research examining how and why experts' needs are misinterpreted by those introducing the new AI technology. Second, outside of studies surveying organisations and looking into higher-level barriers to AI adoption (e.g., environmental, organisational) [1,50], few have explored the emotional, contextual, and collaborative obstacles that might have to be removed before proposing more usable, explainable technologies to domain experts. The AI-related experts' training needs are also underexplored.

Rather than exploring the technology features influencing AI acceptance, we focus on understanding i) the practical and contextual barriers that experts face when AI systems are introduced into their work practices, ii) the AI practitioners' role and their perceptions of the reasons for low AI adoption among experts and iii) how they approached this issue. We then iv) compare practitioners' and experts' perceptions and follow how inaccurate assumptions can stifle effective AI adoption. Lastly, we study v) how effective collaboration between domain experts and practitioners is reflected in experts' motivation to use AI systems. We report on an ethnographically informed study of 10 in-depth interviews with AI developers and natural scientists working at the same organisation, which faces slow adoption of AI systems and bottleneck issues, with more data being collected than possible to process manually.

Our analysis reveals the miscommunications and frustrations between practitioners and experts and how they create barriers to AI adoption. We show that practitioners overestimate the technical knowledge of expert scientists and create an environment in which experts are expected to manage aspects such as software installation and command lines independently. However, the lack of support in performing these simple tasks put a huge emotional burden on experts, took their attention from their projects, and made them resent the AI system. Further, we show that experts are likely to reject an unfamiliar system if its capabilities are not communicated in a relevant way to their project but are willing to invest time in exploring them if they understand how AI aligns with their needs. However, practitioners focus on communicating the technical parameters of their systems, hoping that experts will discover how they can use them in their projects while experimenting with the system. Finally, we report positive cases showing that continuous and goal-driven collaboration can help resolve these miscommunications and develop mutual understanding between teams representing different disciplines.

The paper makes three contributions:

1. We provide AI acceptance requirements that could help to overcome the obstacles to effective AI adoption: i) actively supporting experts during the initial stages of system use, ii) communicating the capabilities of a system in a user-relevant way, and iii) following predefined collaboration rules.
2. We present recommendations for effective collaboration to support the implementation of AI systems in workplaces. These findings could inform practices in various domains and be generalised beyond the case organisation. By following our recommendations, organisations could enable effective AI adoption, scaffold the development of usable technologies, and allow experts to benefit from the systems developed to support them and focus on using and developing their expertise.
3. We describe the emotional and practical consequences of the sudden increased pressure for domain experts to adopt AI tools in their work and the lack of transitional support they receive. We also draw attention to the misappreciation of experts' relatively limited computational skills. We expect these findings to inspire a shift in the research of AI applications in expert contexts.

## 2 RELATED WORK

### 2.1 AI in experts' workflows

AI should support experts by automating time-consuming and mundane tasks and allowing them to focus on exercising their expert skills [82], but it often has the opposite effect [25,57,66]. Researchers studying automation have shown that introducing it has disrupted experts' workflow and changed how they approach their tasks [38], adding to experts' busy workloads and causing frustration [21]. Experts also report feeling a loss of control and agency when relying on automation or AI [37,72,80]. For example, experienced weather forecasters felt that automation made them less adaptive and passive, pigeonholing and disconnecting them from their preferred way of data analysis [69]. Feeling disengaged can force experts to return to more familiar but less effective methods, such as manual information search [43]. Experts often need to change their working methods to accommodate technologies that do not fit their workflows. Interviews with 28 workers facing high job demands using productivity assistant technologies showed that the view of users' work habits presented by the technology did not match their experienced reality. This mismatch required significant time and cognitive efforts for them to adapt [11].

### 2.2 AI adoption

Experts often distrust AI advice, even when it consistently outperforms human judgement (they show algorithmic aversion) [6,15,16]. This effect can be influenced by the specific features of the technology (e.g., explainability [3,8], accuracy metrics [83], perceived power [31]) and various individual and contextual factors, such as level of expertise [44,56]. Several studies suggest that having some control over the AI outcome could mitigate users' algorithmic aversion [17,67,71]. When algorithmic aversion is not addressed, it can lead to decision-makers relying on less effective manual methods and failing to benefit from AI systems [43,78].

Barriers to AI adoption are less researched than algorithmic aversion. Automation studies show that experts are unlikely to rely on automation that lacks transparency, especially if its suggestions are misaligned with their judgement [77], threatens to degrade their skills [42] or jeopardises their agency [4,36,51]. Similar results were observed in AI use cases. For example, medics were unwilling to integrate AI tools into their workflows if they perceived them as threatening their clinical autonomy or if the clinical value of these tools was unclear [29], whereas intelligence workers either reduced frequency or altogether ceased using AI if they observed it underperform [18].

Several studies explored the issues of AI adoption at the organisational level. Alsheibani et al. [1] conducted a questionnaire study involving 207 different-sized organisations about the main barriers to AI adoption. They reported environmental barriers, such as regulations, organisational obstacles related to a lack of top management support, and technical barriers, such as a lack of AI skills and employees' fear of change. Pan et al. [50] studied the challenges of AI adoption in recruitment. The authors emphasised reducing obstacles to AI adoption rather than just developing practical AI techniques for the future of AI technology. They also drew attention to AI practitioners' efforts to understand the difficulty of using AI technologies with little or no technical background. They suggested that reducing complexity and making AI tools user-friendly should be prioritised to increase AI adoption [50]. Factors such as the complexity of the task or organisational impacts are currently underexplored when trying to understand the AI adoption [45].

### 2.3 Theories of technology acceptance

The two most widely accepted theories of the user adoption of technology are the Technology Acceptance Model (TAM) [73,75] and the Unified Theory of Acceptance and Use of Technology (UTAUT) model [39,75]. TAM focuses on the relationship between perceived usefulness and ease of use of new technology, predispositions towards the latest technology and behavioural intentions to use the technology. An individual's decision to engage or not engage in the behaviour of using a new technology is thus based on the expected outcome of using the technology and whether using a new technology is free of effort. The UTAUT model extends the TAM approach by proposing that technology acceptance depends on how widely others accept it by observing others within one's social interactions and experiences [62,75]. We argue that current research efforts exploring AI-expert interactions overlook the aspects of perceived usefulness and required effort. There has been little effort to examine practical hurdles and workflow changes that might prevent experts from using newly introduced AI systems. There is insufficient understanding of the barriers users face before and during the interaction with AI and how they perceive potential future benefits that could outweigh these barriers. In line with Pan and colleagues [50], we argue that the importance of burdens related to domain experts having little or no technical background is underexplored. Research shows that experts understand that they need AI to advance their work [65]. However, there seems to be a hurdle blocking users from adopting and effectively using AI systems. It is essential to understand the factors that impede users before trying to enhance the appeal of the technology.

## 3 METHOD

### 3.1 Procedure and Participants

The research was conducted in a large natural science research centre where the AI-enabled technologies are developed by the in-house team for the scientists within and outside the organisation. The organisation was chosen as a case study because i) AI systems are essential for experts' progress (as the volumes of data are too large to process manually), but the adoption of the AI-enabled software remains low; ii) the advances in data collection technologies are recent, and experts are new in having to rely on AI systems; iii) practitioners and experts work in the same building, and some projects require their collaboration.

We define *Practitioners* as software developers working on developing AI-enabled products and services. We define *Experts* as scientists who work with available technologies to answer scientific questions related to natural sciences. We recruited practitioner and expert participants using a snowball sampling [23]. Our inclusion criteria were participants working in the office and being available for in-person interviews. For expert participants, using AI-enabled software had to be a part of their project. We wanted to understand i) the key barriers preventing users from effectively adopting readily available AI-enabled software that they need for their projects; ii) how the views of experts and practitioners align when reflecting on the same processes, and how that relates to the low adoption of AI-enabled software; and finally, iii) how teams collaborate and what role this collaboration play in improving effective adoption of AI systems.

### 3.2 Data collection

The data collection was guided by a contextual enquiry method [58]. The first author conducted in-depth semi-structured interviews with five practitioners (including the AI team lead and science director) and five expert scientists (in various stages of their PhD projects). During the interviews, experts were also asked to demonstrate how they would use the software. Some practitioners showed relevant material (guides) and software to illustrate their answers, but this was not a planned part of the interviews, and, therefore, the notes from observing this were used only as additional enrichment during the data analysis. Participants signed a consent form before starting the interviews. All interviews were conducted in person, in AI team members' and experts' familiar working environments. The interviewees were not incentivised for their participation. The data sampling was guided by the Grounded theory approach, using Theoretical sampling methods [5]. The interview with the AI lead and science director revealed potential communication issues during the software's introduction and showed that practitioners might have assumptions about why communication does not work and what experts expect—this led to including additional questions in subsequent interviews.

Our interview protocol had three main parts. First, participants were asked about their roles, practices, and workflows to gain background information and context. The second part pertained to their general experiences and attitudes towards using and developing AI-enabled systems. Expert participants were asked how they chose and trusted these systems, how they felt they understood them, how they were first introduced to the new software, and what training they underwent. Practitioners were asked how they develop and introduce usable software to potential users. The third part of the interview was focused on collaboration experiences. Both classes of participants were asked to share experiences and expectations regarding collaboration between practitioners and experts. The interviews were semi-structured and guided by participants' answers.

### 3.3 Data analysis and positionality

Each interview lasted between 45 to 90 minutes. All interviews were conducted in English. The author recorded and transcribed the interviews and analysed notes from the observations. Data was analysed using the Grounded theory approach and advanced coding method [9]. Initial codes were used to fracture the data into categories that were then transformed into more abstract concepts and their dimensions. The relationships were identified between categories and the central core category, creating a storyline supported by explanatory connecting statements [10].

Because of a rich contextual understanding of the data they gained while residing in the organisation, the analysis was conducted by the first author, who collected the data. The author spent four days working in an open office alongside the AI team and experts, where they were part of informal conversations and observations that contextually enriched the collected data. The first author acknowledges that their experiences and positionality could have influenced the perspective and approaches regarding data collection and analysis.

## 4 RESULTS

In this section, we report the categories that emerged during the analysis, the interrelations between these categories and a core category of *assumptions*. We compare the answers of the practitioners and experts to emphasise the miscommunications between the teams and link them to the key frustrations. The letter *P* next to the participant number demotes the practitioners, and *E* refers to the experts. In the interviews, experts referred to AI-enabled systems as software, and this term is used in the following sections.

### 4.1 Using unfamiliar software

In presenting the challenges that experts faced when using new AI systems, we detail four key frictions occurring during the initial stages of use. Each friction is juxtaposed with the practitioners' associated assumptions and illustrates how these misapprehensions could lead to unresolved issues.

*4.1.1 Barrier 1: Not knowing how to get the software to work.*

When experts were asked to demonstrate the software, all observed experts struggled with the command line and had to follow available documentation online. They were frustrated and expressed how stressful the installation and initiation processes were without having programming knowledge: *"[…] it is absolute hell to try to install in the first place. And if you do not have any programming background, you lost."* (E05). Other experts said: *"[…] I find the command line the most difficult thing ever."* (E06). When asked how they know the command line, the experts said: *"Hopefully, someone has told me at some point, and I will have remembered it, but that is not always the case."* (E06).

Despite AI tools being intended to increase the productivity and efficiency of its users, interviewed experts admitted that trying to get the system to work was time-consuming and took their focus away from their projects: *"I have only started my project […] and a lot of it has been trying to get something to work, installation and stuff like that."* (E05) The ones that were still looking for the right software were fearful about how long it will take: *"[…] this is like a major stressor within my project […] I am just going to have to spend ages sifting through this different software, trying to get them to work without having a professional with me who knows how to work it all out."* (E06) They added: *"[…] it is completely taking me outside of what I actually want to achieve with my project because I am having to put so much time and effort into it."* (E06). Another expert said: *"[…] it is always an unreasonable expectation that we are all going to be, you know, pro computer people. I think it is very important that the software is accessible for fairly standard users."* (E03)

When asked if they spend more time dealing with software issues than their project, one of the experts said: "*Yes. Which I have heard is quite normal, especially for scientific software because it is all very specific*." (E05) Not being able to start using the software stifled experts' progress: *"I just have loads of data that is sitting there that have not been [analysed], has a lot of biological information within it, but I have not been able to take any of it."* (E06)

*4.1.2 Practitioners' assumption: "They must know this much."*

Practitioners admitted that computational knowledge is expected: *"You probably need some basic knowledge about Linux and command line and manipulation […]."* (P04). Another practitioner said: *"[…] You need to be able to use a command line, and you need to be able to think about image data which often has various sort of very specific qualities relating to the quality of the image or the resolution."* (P09) The practitioner seemed to be aware that experts might lack the skills but still overestimated them: *"You assume that these days a lot of people have some knowledge of machine learning if they are a scientist […] you assume a certain background, but you are probably often not right."* (P09) Some technical aspects are seen as common sense by practitioners and not something they need to explain to experts: *"I have worked with* *somebody recently who was effectively a user […] and the level of computational knowledge that they had was incredibly limited to the point of things that I do not even consider. It is not even something you think about anymore."* (P02) Tasks that were not directly related to software development (i.e., help with software installation or updating instructions) were seen by practitioners as secondary, something that someone else would do: *"Somebody in a group [scientist team] that wants to use [a software] has the technical expertise and maybe does the job of installing the software, and then other people in the team with less technical expertise use it."* (P09)

*4.1.3 Barrier 2: Incomprehensible, outdated, or missing documentation.*

Experts emphasised the need for easy-to-follow and up-to-date installation instructions and documentation as key features that would encourage them to adopt a software: *"Very clear installation instructions like almost a step by step […] then also step by step what code you should be running and an explanation as to what each of the bits of the code is so that you not have to change it for your specific work."* (E05) Another expert said: *"Easy to install, easy to understand […] having good documentation is probably the main thing."* (E03) The same expert admitted that not having established documentation has been challenging: *"[…]I think that was one of the hardest things I found, just understanding how to use the system and just so many different pieces of software to use. And the documentation, as I said before, there is no established [documentation]."* (E03) Another expert emphasised how important it is that instructions are prepared considering their level of computation knowledge. *"So many of these softwares are just absolutely horrendous to use. […] A lot of the time, there are no instructions online on how to use it, or if there are, they are for people that are very specifically computer science."* (E05).

*4.1.4 Practitioners' assumption: It is available, so it must be useful.*

Practitioners explained that documentation is by default made available to accompany new software: *"[…] we always try to make all our software sustainable and to have good documentation […] so they can visit the documentation at their own time get into deeper end on how to use the software."* (P01) When explaining how users would know how to get a new software working, they explained: *"[…] it has a GitHub page, so it has a repository, and it also has, I think, online documentation, which is quite self-contained, just good."* (P04) When talking about the software they developed, a practitioner said, *"There is the documentation, the manual, and then there is putting an issue on GitHub or sending us an email there was a forum that we had for a while."* (P09) But after being asked if there is anything apart the paper guide for one of their software, the same practitioner said: *"Yeah, it is just that booklet, that kind of thing […] it is likely to be important for them to be able to run [software], especially to install."* (P09) The practitioner referred to the physical guide booklet that one of the interviewed experts tried to use when they could not run the software: *"Luckily, I have remembered that somewhere there is this very handy guide. It is handy because it does explain some levels of how you use [the software]. Whether it tells me how to open it is another thing. Sometimes, I find this guide is just too simplistic. It leaves out a lot of information that I need."* (E06). The expert did not find the necessary information in the booklet.

*4.1.5 Barrier 3. Lack of active support when starting to use the software.*

For experts, asking for help from the in-house AI team was the last option. When asked what they would do if something did not work, they said: *"Ask someone else here [among colleagues] first. There are good forums online […] I have been working with a postdoc here, and he has basically taught me all of this pipeline."* (E03) Another expert said: *"I could give you the easy answer to it - I struggle. […] I think most of the time I try to find material that is out there that maybe other people have made to try and help the users to use the software and things like videos online."* (E06) This expert had to rely on the Internet to find answers: *"[…] I do not have anyone more senior to be able to turn to, to be like, Hey, I see that you have used this software before. Can you help me learn how to use it? […] I am just going to have to work it out by myself. I am just trying to look on the Internet and see if there are any tutorials on there and things like that, which is obviously challenging"* (E06). The expert was emotional about the lack of having somebody actively providing help at the start: "*[…]I know that if someone was around to just teach me how to use it, we could then spend some time, sit down, work it out, practice with using it on some of my data sets. It would probably take a day or two, and then it would simplify the rest of my project […] I am having to struggle through trying to get something to work, which makes you feel stupid sometimes because you are trying to get the software to work and you feel like, Well, why I cannot get this to work? […]It can cause some real existential crises sometimes*." (E06) This expert explained that it is intimidating for a scientist to admit that they do not know something so seemingly basic, so they do not reach out to the AI team with these questions: *"There is definitely an element of pride […] I would say there is an element of not wanting to admit that you do not understand something."* (E06) Another expert also emphasised how challenging it was to find solutions independently: *"[…] I was trying to go through literature as to how maybe people who have helped develop it a little bit were using it to try and figure out what code I was meant to be using. Um, no joy whatsoever."* (E05) Receiving initial support, improved how experts approached the software and how they dealt with issues later: "*[practitioner] was trying to get involved in actually seeing how the code runs and everything and seeing how we were able to run it ourselves eventually. So [practitioner] does not have to every time we want to use something run it for us."* (E10) Getting answers to simple questions led to learning about the software and being more confident and independent in using it: "*[practitioner] has been fantastic at not just telling me what to do but helping me to understand a bit deeper as to what I am supposed to do in the future […] he is helping me so that I am not as lost every single time I use it."* (E05)

*4.1.6 Practitioners' assumption: If they need help, they will ask for it or find it.*

When asked about troubleshooting, one of the practitioners said: "*I have taught programming for quite a few years. We always tell students, Go see your friend, search if it's open source if it has a repository and GitHub [..]*." (P04). This answer also shows that the practitioner compared experts to their programming students, showing a certain level of expectation they have for the expert users, assuming that the same type of material and training methods should suit both groups. When they did have to provide support, frustration was evident: "*I just sent them a very explanatory email of how exactly you have to do it […] what is the correct word about it, customer service or something*." (P01) Practitioners felt frustrated by the way they were being approached by the experts: "*[…] most of the time they are like, I do not know how to do this, please fix it. And there is no clear like, Please, can you do ABC? It is up to us to figure out ABC, and then we work with them again to say well, I manage to do ABC for you. What do you think? And like I actually wanted X, Y, Z.*" (P08) After introducing software, practitioners rarely reach out to experts to find out whether they are able to use it: *"We train them to use the software and then let them get on with it and then the process of working through. You know, it is just a case of emails that backwards and forwards, if they have questions"* (P02). And: *"No, I do not do it [reach out] actively. It is up to them."* (P09).

## 4.2 Effective collaboration and communication as a solution

In this section, we reflect on collaboration between experts and practitioners. We present cases where collaboration led to removing some of the frictions described above, as well as factors that led to successful collaboration or hindered it. Two interviewed experts worked with the AI team as part of their projects and were more positive and involved than other experts. Practitioners also reflected on their experiences of collaborations as successes. These stories involved users becoming advocates for the software, getting involved and finding the confidence to approach each other.

*4.2.1 Collaboration can help establish mutual understanding over time.*

In one project, a group of practitioners and experts had to collaborate on developing the software in weekly meetings. The communication between the teams was not effective in the beginning: *"At the start, we could not communicate with each other at all."* (E10). The expert said that they would simply not understand each other's languages and would have to have someone with both backgrounds translate. *"We were speaking in chemistry, and they were speaking in AI. And I genuinely felt like we were speaking different languages, and the meetings would go round in circles a lot."* (E10). After around three months, they became better at communicating and having a shared goal helped. *"But we have gotten a lot better at it. […] We needed time where we all learned a bit more to be able to talk to each other. I think everyone would agree because you are trying to figure out how to make [software] better."* (E10) One practitioner also reflected that working with experts helped to understand scientists better: *"I find it easier to talk to a biologist now that I have spent a lot of time with them. […] But if you are someone who is completely in your field and all the people you speak to on a daily basis are computer scientists, if you are trying to speak to a biologist, it is a bit difficult to bridge that gap."* (P08)

*4.2.2 Collaboration can help to communicate software capabilities.*

Establishing mutual understanding through collaboration can help to communicate software capabilities to experts. Otherwise, experts might not find the software relevant enough to invest their time: *"[…] I do not want to put a lot of time into learning how to use something when I do not actually know if the software itself is good enough for what I need it to be able to do because that would just be a complete waste of my time."* (E06) A practitioner said that capabilities had not been communicated effectively in the past: *"The scientists did not know what [software] could do. It was not explained very well because the way that it was presented in our science talks and stuff was very much like a computing thing […] the presentations will always like what does the [software] do rather than what does [software] do for you."* (P08). The same practitioner shared a success story of applying this practice: *"We took some data which people were actually working with and showed them what they could do, and they were like, Actually, this is really cool. Like, why did we not know about this before? And we were like, It has always been here. It just did not get across."* (P08) Experts were willing to learn more about the software if they saw how, it could be useful to them: *"[…] when you know that you want to use it, I find it interesting, and I know it is going to give me what I want*." (E10) Experts sought information about the software relevance to their projects online: *"*On the website, they were like […] here is an example of someone using this software for this specific dataset or this specific dataset, you could at least get an idea of whether it might be applicable to yours."(E06)

*4.2.3 Continuous collaboration can help to align expectations.*

Collaborating is essential for effectively aligning expectations and finding middle-ground solutions: *"Mostly what those meetings consist of is all of us talking, right? Does that make sense as a chemist? […] From my point of view, it is a lot of just kind of hearing what [practitioner] is trying and try to make sense of whether that makes sense from the chemistry side of things."* (E10) Effective collaboration was based on maintaining the relationship throughout the development process and frequently checking in with each other: *"There is this continuous communication regarding how the computer science attempts to solve certain issues and if this is okay with them [experts] […] it evolved around what we can do and what we cannot do […] there is a little back and forth to find a middle ground and something that can be done."* (P01)

Continuous communication led practitioners to reflect on their assumptions about users' needs: *"[…] now that we are trying it with them, sitting with them, we are finding that they do things differently than how we expected them to"* (P08)

Practitioners emphasised the importance of communicating throughout the development process as it gives insights into the field and helps to consider details relevant to experts: *"It is very important […] to bring in somebody who is an expert and get some feedback from them because otherwise, you can spend a lot of time just talking about blobs, and you are not talking about what you need to be talking about."* (P09). Another practitioner added that you need to understand the needs of the experts: *"I think it is really important to spend time with people in the domain that you are working with and understand what is important to them before we can help them do anything."* (P08)

*4.2.4 Collaboration increases experts' motivation to be more involved.*

When practitioners had to collaborate with domain experts due to their projects, they actively provided feedback about the software to the AI team. The AI team lead reflected on one of the success stories: *"This user was basically happy to sit and use all the software that did not work […] That enabled us to really drive the software towards something that was practically useful."* (P02) An expert who collaborated with developers was willing to invest time and effort in trying the software and helping to improve it: *"Having played around with it and seeing like what the problems I experience are, it is quite nice coming up with ideas, being like, Okay, if I fix this, this is going to make this so much easier for everyone."* (E05) Collaborating with practitioners when testing the software led experts to be more accepting of software issues and proactive in trying to achieve desired results*: "I run two [analysis], and they were not fantastic […] [AI team] are kind of working on figuring that out at the moment. So, instead of just sitting and doing nothing, I am going to try and see if I can get my [inputs] a bit more specific."* (E10)

*4.2.5 Collaboration makes communication more casual.*

One of the experts shared how the courage to ask grew over time, collaborating with the AI team. *"If I do not get it, I will just be like, What are you talking about? Whereas at the start, I was like, Oh, they expect me to know this. I think it is more of a personal thing. […] I was like, I do not want to waste everyone's time here trying to explain to me. […] I know now they would not have thought it was a waste of time in my mind."* (E10) The expert shared that constantly talking to practitioners made it easy to approach them: *"[…] we talk so much, we spend so much time together, and I would just be like, Yeah, hi, remember me? Um, I have this problem."* (E10) An expert who already had an established connection with practitioners was happy to contact them: *"I would normally email the software developers and be like, Hey, this is for some reason not working or isn't working as well as it should."* (E05) They also gave feedback about errors and persisted with the choice of the software: *"I would much rather contact the development team, mostly because I know them now."* (E05) Without having an established connection, it is difficult to give constructive feedback unless it is encouraged and/or initiated by the AI team: *"[…]it is just always difficult to try and give out feedback in a sort of constructive way"*. (E06)

## 5 DISCUSSION

This study and its analysis develop an actionable strategy for tackling AI adoption issues among domain experts. Our findings have revealed practical barriers to AI adoption at the human level. These transcend the influence of software capabilities (e.g., confidence, explainability). Many of these barriers stem from inaccurate assumptions of users' needs and attitudes. We argue that without eliminating them, experts cannot explore the software and benefit from the features, including explainability. Currently, there is a lack of research trying to understand the aspects unrelated to the AI technical capabilities that stifle the adoption. Our study details the importance of the learning hurdles that users face when new systems are embedded in their workflows. Guiding users through practicalities (e.g., software installation), providing active support, and establishing ongoing collaborative efforts are fundamental to increasing AI uptake responsibly. This means that experts would not only make greater use of AI systems but would also be equipped to use them more meaningfully, with deeper understanding and motivation. Our observations also align with both TAM [73,74] and UTAUT [39,75] models, arguing that the factors influencing technology acceptance are driven by the balance between the perceived benefits and effort required. We showed that this balance can be achieved by clearly communicating the system's capabilities and reducing required effort by providing initial and ongoing collaborative support. The following subsections expand on the observed barriers and recommendations to overcome them.

**5.1 Barriers to AI adoptions during the introductory period**

Experts struggled to bypass the initial steps of installing and running the software. They were more likely to reject the system if they experienced issues during this initial stage, especially if the software did not have comprehensible documentation and they did not receive support. Experts complained that practitioners expected them to *be programmers* and to *just know*, for example, what command line instructions to use. However, they did not seek help because, as scientists, they felt embarrassed to struggle with basic tasks. Practitioners, indeed, did not consider that experts might not have these basic skills. They saw tasks, such as software installation, as a matter of fact. When introducing their software to potential users, they used one-off training workshops, predominantly focused on explaining higher-level details, such as underlying mechanisms or the data processing pipeline and then processing some data. Although documentation and contact information were usually available, practitioners did not reflect on whether the available training or documentation was helpful. There was a lack of active support or follow up on experts' progress during the initial stage.

*5.1.1 Recommendation: Offer active support during the initial stages.*

We recommend providing support during the initial stage of AI adoption. Before the new technology is fully introduced to the experts' workflow, they should be given time to adjust to the new system. During this stage, experts should receive active support from assigned team members, who would guide them and help install and run the new system. This should also establish contact between an expert and a person/team to whom they could communicate their questions. Experts should be guided to the updated and comprehensible documentation, which would be established in collaboration with them. Providing support when implementing new technologies could positively influence user attitudes towards technology and their self-efficacy [41]. Experts should actively seek practitioners' feedback and clearly communicate theirs during this stage. During this stage, communication between experts and practitioners could improve how technology fits within experts' workflow [49]. It could also reduce the stress of adjusting to the workflow-related changes. Park and colleagues [52] showed that discussing practical solutions to their AI problems with the management team provided emotional support, lowered social burdens, and demonstrated that managers cared for them. Experts should also be allowed to practice using the new system. Practice sessions can enable a learning-before-doing approach, enhance later performance [55], and increase motivation to use and experiment with the new technology independently [19]. Allowing users to test and reflect on the new technology has also increased its acceptance. A study exploring the learning difficulties of novice technology users revealed that they mostly valued the opportunities to experiment and explore the system freely and safely [7]. An ethnographic study of technology implementation in different cardiac surgery departments reported that sites that implemented trials for technology use followed by reflection sessions improved the chances of successful implementation and formed a learning cycle [19].

## 5.2 Barriers to investing time to learn and explore AI systems

Practitioners argued that users were unwilling to participate in software design and development stages and just wanted a final result or expected AI to *just do its magic.* This felt unfair to practitioners: "*[…] we are expected to learn biology, but they [experts] are never expected to learn the computing side of things*". Experts were actually willing to invest time and effort in learning and improving the software if they knew the system would likely be helpful for their project. Experts prioritised systems that had any evidence of being beneficial for their specific project. For example, they searched for videos on Twitter, with scientists demonstrating software and explaining its features in the language and context understandable to them. Otherwise, they felt they were risking wasting their time. However, the capabilities of the systems were not communicated effectively to them. Practitioners would present the technical aspects of the software without framing its capabilities in a way that would be relevant to experts. Practitioners also assumed that experts would discover software benefits independently and then communicate them to their peers.

*5.2.1 Recommendation: Clearly communicate the capabilities of a new system.*

When introducing a new AI system, clearly communicate how experts could benefit from it. Communicate not what the system can do but what the system can do for them as experts and for their projects. Use domain-specific language and avoid technical jargon. Research in digital skill education showed that improving users' involvement and facilitating their learning requires the identification of specific functions of the technology for each member based on their expertise [13]. Our interviewed experts were willing to invest their time and effort if they could see how they could benefit from the system. This aligns with research showing that the relevance of the learning material and tasks is an important motivator to seek more knowledge [35].

Showcase the capabilities of the system by using practical demonstrations. One of the most convincing factors for the adoption of a new system was seeing other experts use it on a similar dataset or project. For example, one of the experts decided if they would invest time in trying the software and learning about it if they saw another scientist demonstrating it on Twitter. We recommend including practical demonstrations (not for teaching purposes) tailored to the experts during training workshops. If this is not possible, we recommend preparing short videos in collaboration with domain experts that would be made available online or internally to show how it could be used and benefit them. Begin talking about the potential capabilities of the software with experts before it is fully developed. It can help to see if experts' and practitioners' expectations align, better understand experts' needs, and avoid developing unusable systems. It could also help experts feel more in control and less averse to the technology [17,64].

## 5.3 Barriers to AI adoption due to effective collaboration between experts and practitioners

Our study revealed the importance of ongoing collaborative efforts for responsible AI adoption. Infrequent communication without a pre-defined plan and set meetings did not lead to established collaboration, even when teams worked in the same building. Without it, members of both groups were increasingly frustrated because of the expectations projected onto them. Our results showed that consistent and planned collaboration could resolve these frustrations. When experts and practitioners had to work together, they established mutual understanding, aligned their expectations, and learned how to ask questions and explain information that would be understandable and meaningful to the other party. After collaboration, experts were motivated to learn and troubleshoot independently and invest time in taking steps to improve the software. When two teams worked together because of the nature of the projects, experts actively got involved, and even when the software did not perform as they expected immediately, they knew that it was still in the stage of development. Instead of rejecting it, they gave experts feedback that could help improve the system. Expert engagement can help define expectations for system performance and identify contextual factors that are likely to change when the system is implemented in practice [18].

*5.3.1 Recommendation: Plan the collaboration.*

Implement the plan for collaboration between the experts and practitioners (or other stakeholders). For example, start with agreeing on the type (online, in-person), the frequency and the minimum timeframe for collaboration. The collaboration does not have to be extensive in the time spent working together, but it must be planned and continuous. Having collaboration planned could take the burden off someone having to initiate meetings. It could also change the dynamic from experts only contacting practitioners to reporting problems or at the stage where they already feel frustrated. It could also help to continue when collaborative efforts are not immediately beneficial. Our results showed that establishing an understanding between multidisciplinary teams required continuing collaboration even when it felt ineffective. Poorly planned practices have been shown to complicate effective collaboration [63]. Having a collaboration plan improved engagement from both teams. Improving interdisciplinary communication skills is essential for effectively providing and receiving feedback [27,85]. Effective user feedback is necessary for generating, refining, and fine-tuning design options and alternatives [27]. Feedback is critical to framing arguments and balancing multiple perspectives and considerations in the design process [27]. Collaboration practices, such as team discussions, should also be applied during the initial stages of system use with the goal of learning. Learning through discussion is an effective way of acquiring new information. Collaborating can create a dynamic learning system with humans in the loop [70].

Define collaborative goals for different stages of the process. The ability to work effectively in teams is driven by having a common goal and assuming shared responsibility for completing tasks [48]. For example, during the planning or design stages, collaborate on aligning expectations and trying to understand experts and their workflows better. Our study showed that starting collaborative efforts before the software was developed helped meet users' needs better. Early efforts also made users more confident, independent, and persistent when testing the new system, troubleshooting it, and communicating issues and feedback. Collaboration could also be applied in the mid-stages of software development, with a goal to demonstrate and discuss a preliminary version of the project. Involving users through collaboration is necessary for effective AI development [32]. Our interviewed expert, involved in a collaboration, volunteered to test a Beta version of a new product. The argument that experts' needs should guide the development of the technology is in line with user-centred HCI research goals [28]. Including users in the design process is also the basis of the participatory design methods [86], which have been effective in various domains [2,61].

## 5.4 Further Implications

Our study revealed that the growing pressure and necessity to use AI systems can put a significant emotional burden on experts. Introducing new technologies to experts' workflows without appropriate support or preparation negatively affected their emotional well-being and reduced the time they could spend on their projects. The interviewed scientists were experts who had access to all the best equipment, creating the potential for making ground-breaking scientific discoveries in life sciences. However, they were spending a lot of time figuring out how to use software, unable to process available data. Experts admitted that this process affected their wellbeing, quality of work and confidence. Addressing our recognised issues could reduce this stress, allowing experts to focus on their expert tasks and progress with their work. Our study also emphasised an issue of lack of appreciation for experts' support needs with practical aspects, such as command lines and installation processes. Preparation and ongoing assistance should be prioritised when introducing new AI systems.

Otherwise, there is a risk that the application of expert skills will be limited across domains that could greatly benefit from adopting AI.

Applying our suggestions could help AI developers calibrate experts' expectations, motivate them to learn about the AI system, advocate it to their peers, effectively communicate their feedback, and be more independent users. The interviewed practitioners referred to the lack of resources, such as time, that prevented them from providing more support for experts. However, practitioners admitted that their approach was ineffective, focusing too much on the technology and presenting it to experts only when it was developed. This led to cases where their software was unused and had to be abandoned. We suggest that initial time can be found by reprioritising tasks and that it will pay off later, as practitioners would have to spend less time troubleshooting and advocating for their products. Following our steps towards better collaboration could also facilitate research approaches, such as participatory design, requiring meaningful user engagement and collaborative efforts between stakeholders.

## 5.5 Limitations and Future Directions

Our study focused on a specific domain and two teams working within the same organisation and environment. A study involving more participants with varying levels of task and domain expertise is necessary to explore further the practical barriers to AI adoption arising at various stages. These findings provide meaningful insights regarding AI adoption among experts that could be generalised across different areas and domains. However, further studies are needed to test our recommendations in domains of varying risk, time pressure, and the extent to which human experts rely on AI.

This study revealed that the practitioners' lack of appreciation for experts' limited computational knowledge prevented experts from effectively adopting AI tools. Experts often do not have a choice but to use AI to exercise their expertise. The pressure to use new technologies will only increase with further advances in AI and data collection methods. Researchers should think of ways to support and motivate experts to learn about AI. But they should also explore ways to educate developers to communicate information about their developed systems effectively and support users more proactively. Future studies should examine ways to support mutual learning processes through collaboration.

## 6 CONCLUSION

Research analysing AI adoption often explores users' actions and feelings towards AI outputs or how certain software features affect users' trust and reliance on AI. The practical hurdles users face when these systems are embedded in their workflows are significantly less explored. Our study has revealed that guiding users through mundane practicalities (e.g., software installation), providing active support, and establishing ongoing collaborative efforts are fundamental to increasing AI uptake. The paper has presented steps likely to help remove the key barriers separating experts from the systems they need and want to use. We propose i) actively supporting experts during the initial stages of software use, ii) clearly communicating the software's capabilities in a way that is relevant to the experts' language, and iii) following predefined collaboration rules. We argue that following these recommendations could make experts more accepting of AI systems and increase their motivation to understand the technology they use, advocate it to their peers, and be more independent and confident using it. Moreover, it could help them to be more specific in communicating their feedback to practitioners. Overall, supporting experts more effectively and promoting between-team collaboration could make them more conscious and responsible AI users, allowing them to benefit more from new technologies. Our findings could facilitate research approaches, such as participatory design, requiring meaningful user engagement and stakeholder collaborative efforts to improve AI technology's contextual fit and usability.